\documentclass[runningheads]{llncs}
\usepackage{graphicx}
\usepackage{amsmath}
\usepackage{amsfonts}
\usepackage{multirow}
\usepackage{float}
\usepackage{multirow}
\usepackage{color}
\usepackage[colorlinks]{hyperref}
\usepackage{subcaption}
\begin{document}
%
\title{Recovering medical images from CT film photos}
%
%
\author{Quan Quan\inst{1} \and
	Qiyuan Wang\inst{2} \and
	Yuanqi Du\inst{3} \and 
	Liu Li\inst{4} \and
S. Kevin Zhou\inst{1, 2}
}

\institute{\
Key Lab of Intelligent Information Processing of Chinese Academy of Sciences (CAS), Institute of Computing Technology, CAS, Beijing 100190, China \and
Medical Imaging, Robotics, and Analytic Computing Laboratory and Engineering (MIRACLE) 
School of Biomedical Engineering \& Suzhou Institute for Advanced Research, University of Science and Technology of China, Suzhou 215123, China \and
George Mason University \and
Imperial College London 
}

\authorrunning{Q. Quan, et al.}

\maketitle       
\begin{abstract}
While medical images such as computed tomography (CT) are stored in DICOM format in hospital PACS, it is still quite routine in many countries to print a film as a transferable medium for the purposes of self-storage and secondary consultation. Also, with the ubiquitousness of mobile phone cameras, it is quite common to take pictures of CT films, which unfortunately suffer from geometric deformation and illumination variation. In this work, we study the problem of recovering a CT film, which marks \textbf{the first attempt} in the literature, to the best of our knowledge. We start with building a large-scale head CT film database CTFilm20K, consisting of approximately 20,000 pictures, using the widely used computer graphics software Blender. We also record all accompanying information related to the geometric deformation (such as 3D coordinate, depth, normal, and UV maps) and illumination variation (such as albedo map). Then we propose a deep framework called \textbf{F}ilm \textbf{I}mage \textbf{Re}covery \textbf{Net}work (\textbf{FIReNet}) to tackle geometric deformation and illumination variation using the multiple maps extracted from the CT films to collaboratively guide the recovery process. Finally, we convert the dewarped images to DICOM files with our cascade model for further analysis such as radiomics feature extraction. Extensive experiments demonstrate the superiority of our approach over the previous approaches. We plan to open source the simulated images and deep models for promoting the research on CT film image analysis.
\end{abstract}

\begin{figure}[t] 
\centering
\begin{subfigure}{.45\textwidth}
\centering
  \includegraphics[width=\linewidth]{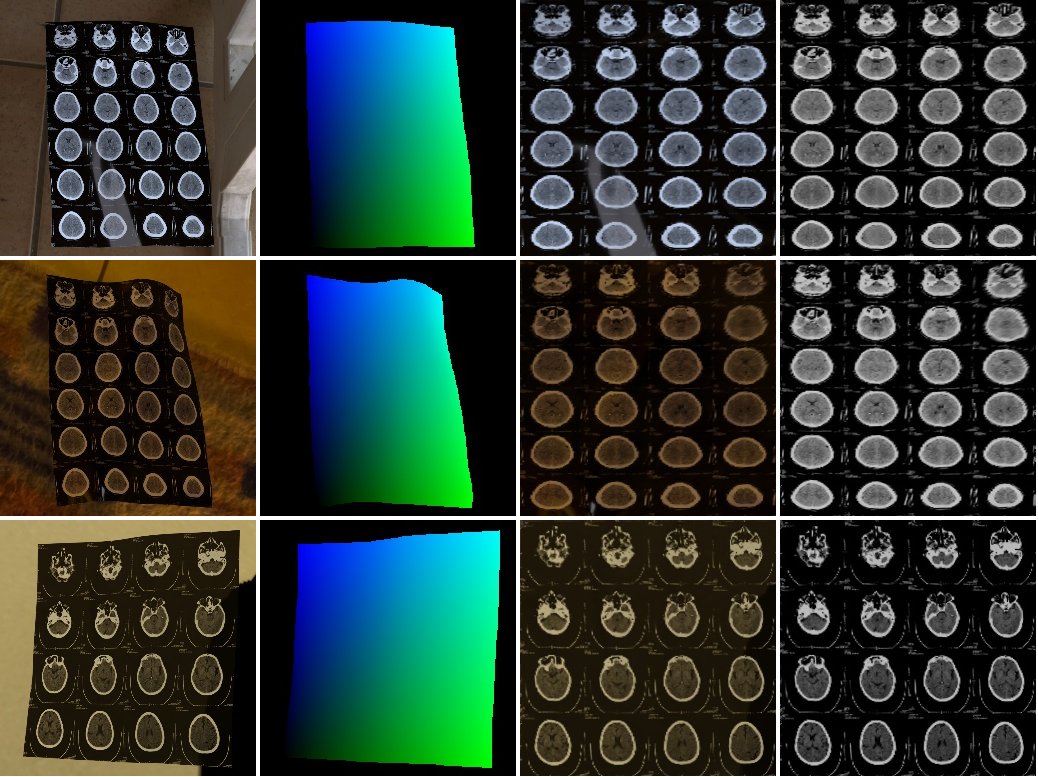}
  \caption{}
  \label{fig:demo1}
\end{subfigure}
\begin{subfigure}{.45\textwidth}
\centering
  \includegraphics[width=\linewidth]{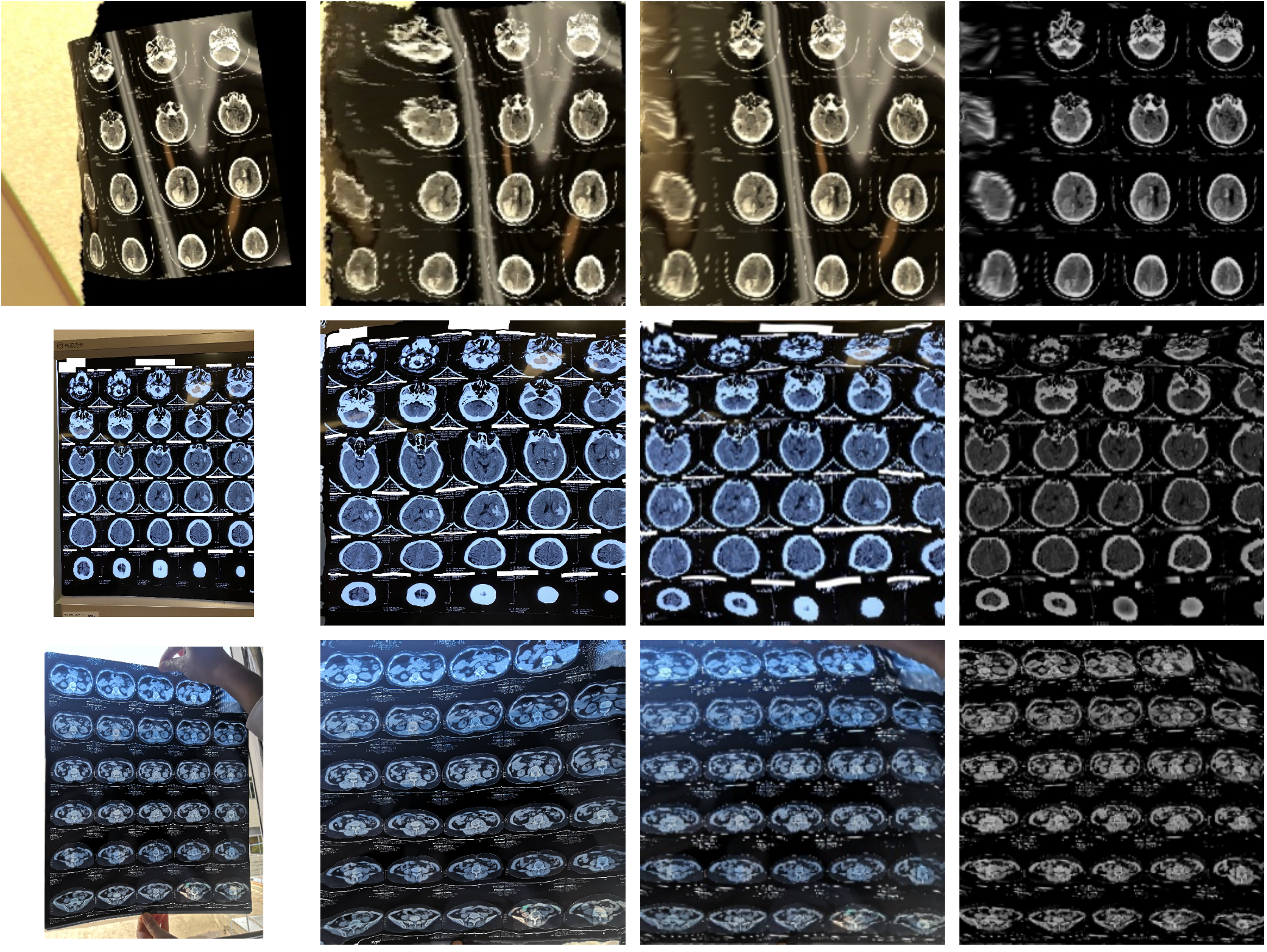}
  \caption{}
  \label{fig:visual_comparison}
\end{subfigure}
  \caption{(a) \textbf{CT film recovery.} The 1st column: warped CT films. The 2nd column: UV map. The 3rd column: dewarped films. The 4th column: de-illuminated \& dewarped films. (b) \textbf{Qualitative comparison between our approach and others.} The 1st column: warped CT films. The 2nd column: DewarpNet~\cite{das2019dewarpnet}. The 3rd column: ours. The 4th column: de-illuminated images. The 1st row: difficult simulated photo. The 2nd and 3rd row: real photos}
\end{figure}

%

\section{Introduction}
 Recently, more and more attention has been drawn to medical image analysis. 
 While medical images such as CT are stored in DICOM format in hospital PACS, it is still quite routine in many countries to print a film as a transferable medium for the purposes of self-storage and secondary consultation. For example, the market size of global medical radiography film, which is commonly used to print a stack of CT slices, is estimated to reach 986.1 million USD by 2026, at a CAGR of 0.4\% since 2016\footnote{https://www.marketwatch.com/press-release/medical-x-ray-film-market-size-share-global-trends-market-demand-industry-analysis-growth-opportunities-and-forecast-2026-2020-11-03}. 

With the ubiquitousness of mobile phone cameras, it becomes quite common to take pictures of the CT films to avoid physically transferring CT films or support remote access. 
While convenient, the pictures most time do not meet the standard for the doctors to process and analyze because sometimes the CT films are warped in the pictures.
In addition to the geometric deformation, there is illumination variation caused by different the lighting condition and background scene. Therefore, it is a challenge to both dewarp the pictures and recover the content from rectified CT films in a framework. 
In this paper, we make the first attempt in the literature, to the best of our knowledge, to propose a CT film recovery solution. Its crucial part is to remove the background in the image and rectify it as in Fig.\ref{fig:demo1}.

Although there are no existing works about recovering warped CT films previously, a more related problem, the recovery and de-distortion of document images, has been studied. The previous methods usually rely on the geometric properties of papers~\cite{meng2018exploiting,zhang2009unified}, 3D shape~\cite{brown2001document,tsoi2007multi,koo2009composition,meng2014active,ulges2004document,yamashita2004shape,ostlund2012laplacian,you2017multiview} or visual cues~\cite{courteille2007shape,courteille2007shape,zhang2009unified,ezaki2005dewarping,ulges2005document,lu2006document,meng2014active,kim2015document,liu2015restoring,cao2003cylindrical,liang2008geometric,tian2011rectification,witkin1981recovering,malik1997computing,forsyth2001shape,das2017common}.
However, these methods are resource- and time-intensive. Later, with the advancement of deep learning on computer vision, researchers start to tackle this problem utilizing the convolutional neural network~\cite{ronneberger2015u}. Recently, DocUNet~\cite{ma2018docunet} firstly tackle this problem by constructing and training large 2D document datasets. DewarpNet~\cite{das2019dewarpnet} follows ~\cite{ma2018docunet} to construct a 3D document dataset and uses the 3D coordinate map, which contains more information about the spatial structure of the documents, to generate a backward map, which is used to restore the warped image in the final step. However, in our CT film recovery task, as in Fig.\ref{fig:visual_comparison}, the backward map will cause many locally-warped areas, leading to poor visual performance.
These locally-warped regions can negatively affect our further medical image processing and clinical analysis for CT films, hence we find another way to obtain the dewarped image\textemdash using a UV map. Though a UV map is generally applied to map textures to models, it can be also used to obtain textures from \textquotedblleft CT film models \textquotedblright, and is verified to generate less locally-warped regions.  
Another problem is that the ambient light mixed with various colors reduce the quality of CT film images, and is harmful to the accurate diagnosis. Besides, RBG images have smaller value range and less details than DICOM files. Therefore, we need to remove the disturbance of lights and restore the value range and detail of the radiographs. 

We propose a \textbf{F}ilm \textbf{I}mage \textbf{Re}covery \textbf{Net}work (\textbf{FIReNet}) that recovers CT film by the UV map.
The FIReNet consists of two parts: 1) \textbf{Dewarping module} includes two modules: a) \textbf{Multi-map module} is designed to improve the quality of UV maps by outputting and combining multiple annotation maps. b) \textbf{Transformation module} is designed to generate UV maps, and also deformation maps for enhancing the UV map. In addition, to focus on disentangling local warps, both maps are trained with additional customized loss. 2) \textbf{Quality Restoration Module} includes two modules: a) \textbf{De-illumination module} is designed to solve the illumination problem by outputting albedo maps which retain only the content of the film without the light information. b) \textbf{Medical Image restoration module} is designed to convert the value range from RGB to CT, and predicts more details ignored in photos. We conduct our experiment based on the below CTFilm20K dataset.

\section{CTFilm20K}
\begin{figure}[t]
\centering
\begin{subfigure}{.5\textwidth}
  \centering
    \includegraphics[width=1\linewidth]{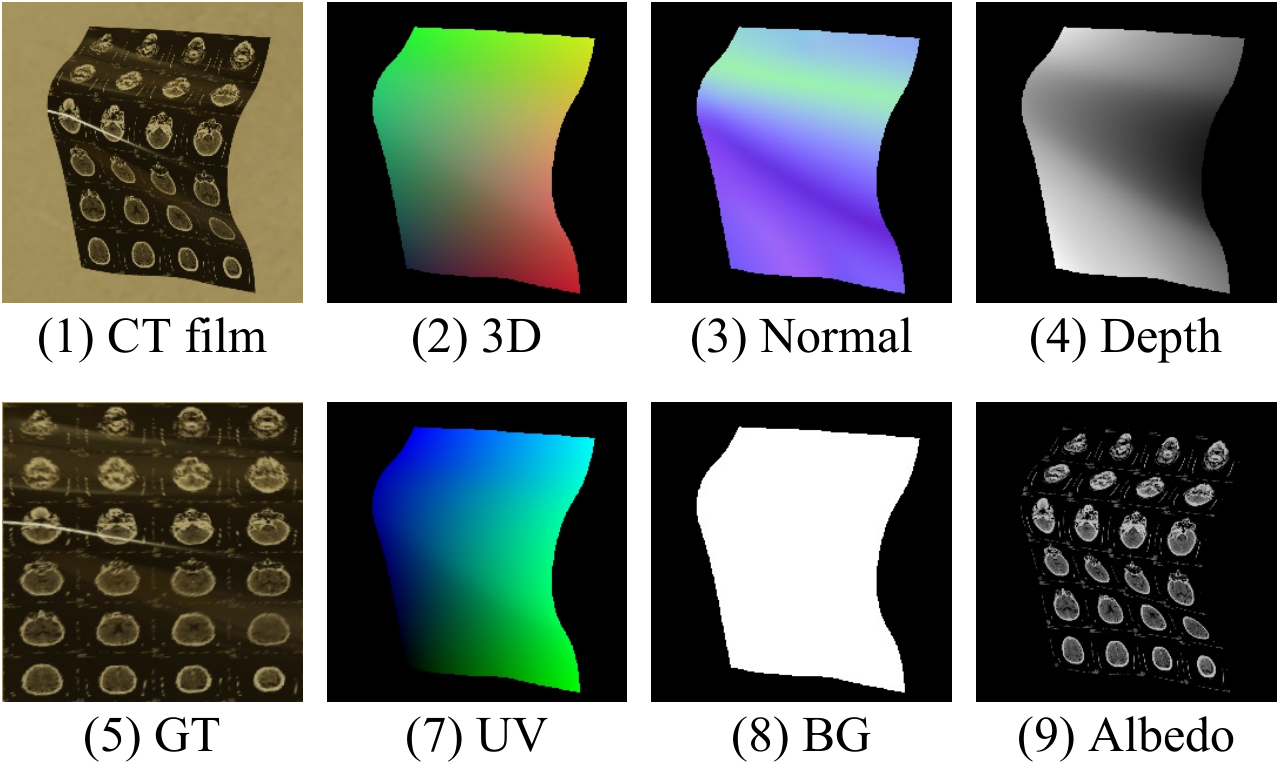}
    \caption{}
    \label{fig:dataset1}
\end{subfigure}
\begin{subfigure}{.4\textwidth}
  \centering
    \includegraphics[width=1\linewidth]{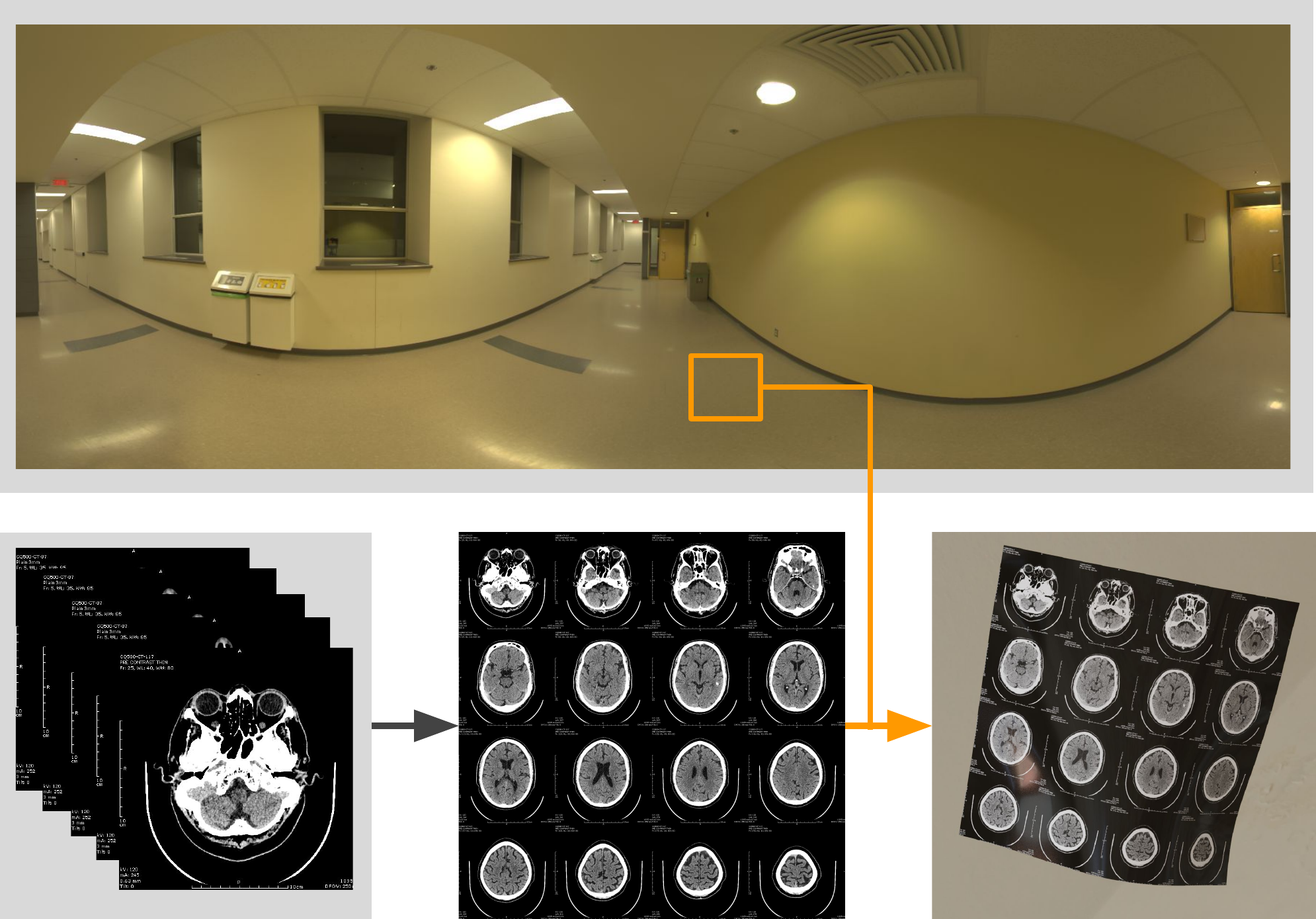}
  \caption{}
  \label{fig:dataset2}
\end{subfigure}
\caption{(a) \textbf{Components in CT film dataset.} (1) Warped CT film. (2) 3D coordinate map. (3) Normal map. (4) Depth map. (5) Dewarped film. (6) UV map. (7) Background mask. (8) Albedo map. (b) \textbf{CT film data construction.} Top: environment images. Bottom-left: a series of CT slices. Bottom-middle: CT films from joint slices. Bottom-right: warped CT films. Black arrow: arranging CT slices to a CT film. Yellow arrow: rendering warped CT films with Blender\cite{blender}}
\label{fig:dataset}
\end{figure}

As three are no public data for CT film recovery, we construct our \textbf{CTFilm20K Dataset}, which contains a large number of CT films with various real-world warping scenarios and different contents. After collecting high-quality background images and CT data, we utilize photorealistic rendering to generate images. Each sample contains several labeled maps including 3D coordinate, normal, UV, and albedo maps (Fig.\ref{fig:dataset1}). In total, the dataset contains about 20,000 richly annotated high-quality film images.

As in Fig.\ref{fig:dataset2}, 
we first extract slices from the public medical CT DICOMs~\cite{chilamkurthy2018development} and convert them in a grid format to generate the film texture. In comparison with Doc3D~\cite{das2019dewarpnet}, which captures the 3D shape of naturally deformed real objects, we rely on the embedded physics computing engines in Blender~\cite{blender}, a rendering software, which simulates deformation for different situations. After that, we render the images with synthetic film textures laid in various HDR scenes from~\cite{DBLP:journals/tog/GardnerSYSGGL17} and HDR Haven\footnote{HDRIHAVEN: https://hdrihaven.com/ \label{hdrhaven}} using path tracing. We randomly set camera positions in a certain range 
and vary illumination conditions in rendering.

\section{Method}

\begin{figure*}[t] 
\centerline{\includegraphics[width=0.85\linewidth]{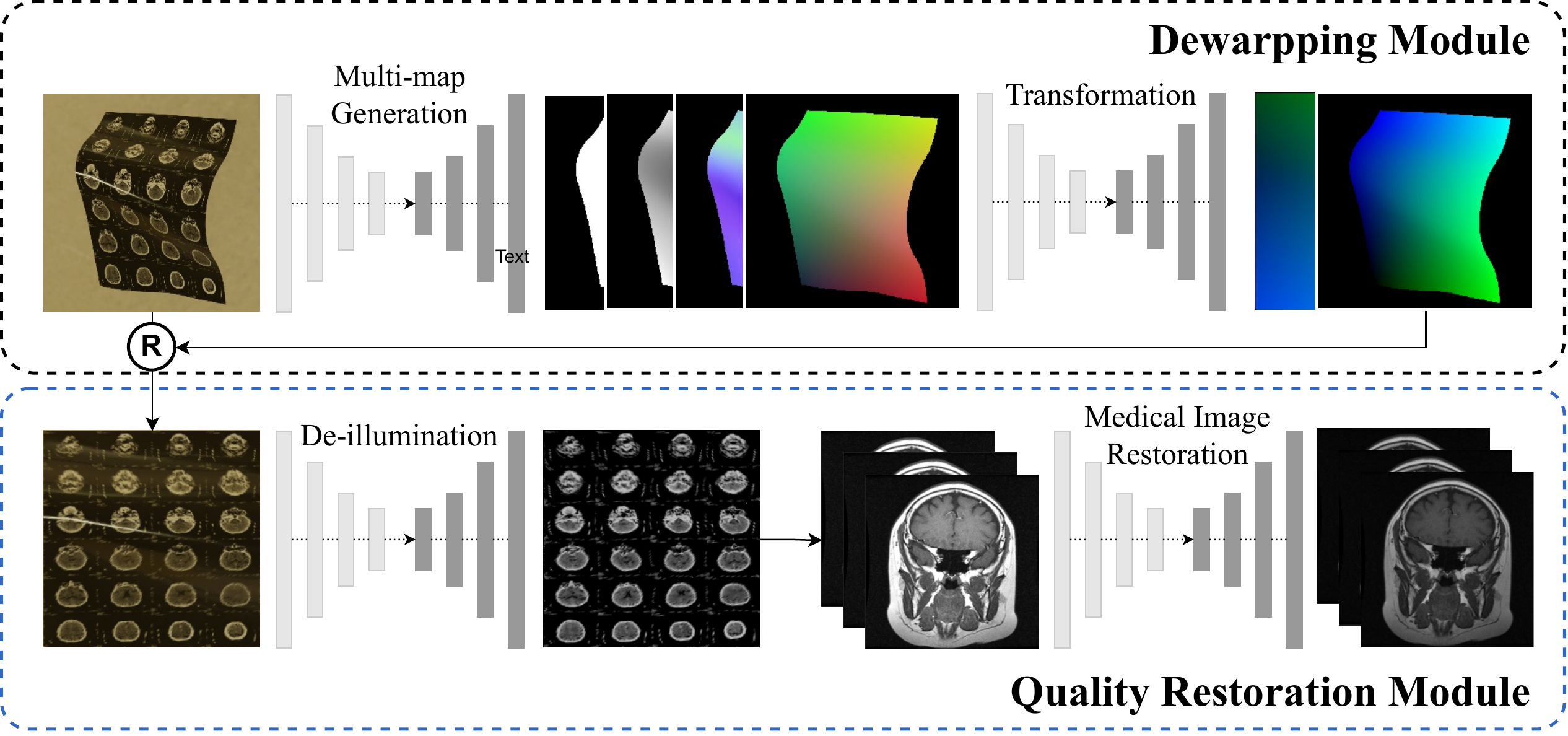}}
\caption{\textbf{Overview of the proposed network: FIReNet.}}
\label{fig:overview}
\end{figure*}

As in Fig.~\ref{fig:overview}, our network consists of two modules: dewarping module and content recovery module. In general, FIReNet takes a warped CT film $\mathbf{I} \in \mathbb{R}^{h \times w \times 3}$ ($h$ and $w$ are image height and width, respectively) as input and predicts a UV map $\mathbf{M_{uv}} \in \mathbb{R}^{h \times w \times 2}$, where each point indicates its position on a texture map. To achieve the purpose of obtaining the texture map, we transform UV map $\mathbf{M_{uv}}$, which maps textures to objects, to the backward map $\mathbf{B}  \in \mathbb{R}^{h \times w \times 2}$, which maps objects to textures by linear interpolation. 
In addition, to satisfy the need of radiologists, we convert the dewarped images to DICOM file, which is used to compute radiomics features. 
\subsection{Dewarping module}

\textbf{Multi-map module.}
We believe that the original image and its various annotations such as 3D coordinate map $\mathbf{M_{3D}} \in \mathbb{R}^{h \times w \times 3}$, normal map $\mathbf{M_{nor}}\in \mathbb{R}^{h \times w \times 3}$, and depth map $\mathbf{M_{dp}}\in \mathbb{R}^{h \times w \times 1}$ are closely related, hence unlike DewarpNet~\cite{das2019dewarpnet} using only 3D map, we use the information of three maps mentioned above by a shared encoder which extracts the features of the input image and 4 different decoders which generate $\mathbf{M_{3D}}$, $\mathbf{M_{nor}}$, $\mathbf{M_{dp}}$ and background map $\mathbf{M_{bg}} \in \mathbb{R}^{h \times w \times 1}$ respectively, with U-Net~\cite{ronneberger2015u} as the backbone network. When these maps are generated, they are combined and filtered by $\mathbf{M_{bg}}$ and sent to the subsequent modules for more processing.

\textbf{Transformation module.}
Instead of generating a backward map $\mathbf{B}$ directly from a model~\cite{das2019dewarpnet}, we design the transformation module to generate the UV map $\mathbf{M_{uv}}$ which is pixel-wise aligned with the original image and is stable and robust for restoration, but some regions especially around the edges are often missed because the number of valid points in $\mathbf{M_{uv}}$ is less than that in the backward map $\mathbf{B}$ and most of the valid points in $\mathbf{M_{uv}}$ are concentrated in the middle part. Therefore, 
we also generate a \underline{deformation map} $\mathbf{M_{df}}  \in \mathbb{R}^{h \times w \times 2}$, which is the relative movement position of each point from the input image to the unwarped image, and we utilize $\mathbf{M_{df}}$ to enhance $\mathbf{M_{uv}}$. Finally, we will use interpolation to make the final image. We also invoke the U-Net as the basic framework and take the integrated features from the previous module as inputs, and then use two different decoders to generate the corresponding UV map $\mathbf{M_{uv}}$ and deformation map $\mathbf{M_{df}}$, both of which have the same size as the original image.
Afterward, we combine the UV map $\mathbf{M_{uv}}$ and deformation map $\mathbf{M_{df}}$ by converting the deformation map to UV map $\mathbf{M^{'}_{uv}}$ where points are used to fill blanks in $\mathbf{M_{uv}}$, which is finally converted to backward map $\mathbf{B}$ for CT film recovery. We follow \cite{das2019dewarpnet} to recover the warped images with $\mathbf{B}$.

\subsection{Quality restoration module}
For de-illumination, inspired by DMPHN~\cite{zhang2019deep,mehri2021mprnet} designed for denoising, we adopt the cascade network as our base model to meet our requirement of large scale, content consistency and clarity. In detail, the dewarped image will be sent to the de-illumination module to get the de-illuminated image $\mathbf{I}_d$. We train this module with noisy images/ground-truth pairs following settings in \cite{mehri2021mprnet}. 
In addition, 
we apply another cascade module to better restore the medical images $\mathbf{I}_r$ with CT values.

Comparison between the predicted image and ground-truth in visual level is often insufficient. To this, besides the common evaluation metrics like SSIM and PSNR, we perform a comparison based on radiomics features. We leverage the statistical tool, Chi-square Analysis, to assess the similarity of two distributions of multiple radiomics features including first-order features like energy, entropy, standard variation, etc. and shape features, and radiomics features like GLCM, GLDM, etc.

\subsection{Loss function}
In our training of dewarping module, we perform the end-to-end training with multiple objectives. In the multi-map module, we train the four maps with \textit{$\mathcal{L}_1$ loss}. 
For example, $\mathcal{L}_{3D} = || \mathbf{\hat{M}_{3D}} - \mathbf{M_{3D}} ||_{1}$, 
where $\mathbf{\hat{M}_{3D}}$ represents the predicted 3D coordinate map, $\mathbf{M_{3D}}$ is the ground truth, and $\mathcal{L}_{3D}$ represents the loss value on $\mathbf{M_{3D}}$. Similarly, We define other losses of $\mathcal{L}_{bg}$, $\mathcal{L}_{nor}$, and $\mathcal{L}_{dp}$. Therefore, the total loss for shape module, $\mathcal{L}_{shape}$, is formulated as follows:
\begin{equation}
\mathcal{L}_{shape} = \mathcal{L}_{3D} + \mathcal{L}_{nor} + \mathcal{L}_{dp} + \mathcal{L}_{bg}
\end{equation}

In transformation module, we first perform the same operations as before on the maps of $\mathbf{M_{uv}}$, and $\mathbf{M_{df}}$ to get the corresponding losses, which are denoted as $\mathcal{L}_{uv}$, and $\mathcal{L}_{df}$, respectively. In addition, we also add a \textit{shifting loss} $\mathcal{L}_{shift}$ to reduce shifting, a \textit{disturbance loss} $\mathcal{L}_{disturb}$ to reduce total variance, and a \textit{de-shifted difference loss} $\mathcal{L}_{diff}$ to reduce variance of each point of $\mathbf{M_{df}}$. Denoting the predicted deformation map by $\mathbf{\hat{M}_{df}}$ and the ground truth by $\mathbf{M_{df}}$, we represent the entire process as follows:
\begin{equation}
    \begin{split}
&\mathcal{L}_{shift}  = ||\mathbf{\mu}||_{1}, ~\mathcal{L}_{disturb}  = ||\mathbf{\sigma}||_{1}, \\
&\mathcal{L}^{'}_{diff}  = min(||\mathbf{\Delta M_{df}}||_{1} , ||\mathbf{\Delta M_{df}} - \mathbf{\mu}||_{1})), \\
&\mathcal{L}_{diff} = \begin{cases}
\mathcal{L}^{'}_{diff}&\mathbf{\Delta M_{df}} \odot (\mathbf{\Delta M_{df}} - \mu) > 0, \\
0 &\mathbf{\Delta M_{df}} \odot (\mathbf{\Delta M_{df}} - \mu) <= 0. \\
          \end{cases}
    \end{split}
\label{eq:transLoss}
\end{equation}
where $\mathbf{\Delta M_{df}}  = \mathbf{\hat{M}_{df}} - \mathbf{M_{df}}$; $\mathbf{\mu} = \mu(\mathbf{\Delta M_{df}})$; $\mathbf{\sigma} = \sigma(\mathbf{\Delta M_{df}})$.

For deformation map, we hope our model to focus on relative deformation, hence in (\ref{eq:transLoss}) we first calculate the difference $\mathbf{\Delta M_{df}}$ between the predicted deformation map $\mathbf{\hat{M}_{df}}$ and ground truth $\mathbf{M_{df}}$, and then calculate the mean $\mu$ and the variance $\sigma$ of $\mathbf{\Delta M_{df}}$. To focus on the relative image warping, we separate the shift loss $\mathcal{L}_{shift}$ from $\mathbf{\Delta M_{df}}$, and add standard variation as disturbing loss $\mathcal{L}_{disturb}$. In particular, we hope the prediction $\mathbf{\hat{M}_{df}}$ to approximate the relative location if far away from both relative and the absolute location, but we should not affect points already close to the absolute location, hence we calculate $\mathcal{L}_{diff}$ as mentioned above.
Therefore, the total loss of the deformation map $\mathcal{L}_{df} = \mathcal{L}_{shift} + \mathcal{L}_{disturb} + \mathcal{L}_{diff}$. 
Similarly, we can get $\mathcal{L}_{uv}$ for UV map according to $\mathcal{L}_{df}$ according to $\mathcal{L}_{3D}$. The loss for transformation module can be represented as follows: $\mathcal{L}_{trans} = \mathcal{L}_{df} + \mathcal{L}_{uv}$. 
Finally, the total loss for training the dewarping module is given as : $\mathcal{L}_{dewarp} = \mathcal{L}_{shape} + \mathcal{L}_{trans}$.

For quality restoration module, we follow \cite{mehri2021mprnet} to perform a end-to-end training respectively. The objective is given as: $\mathcal{L}_{recover} = \frac{1}{2} \sum^N_{i=1} || \mathbf{S}_i - \mathbf{G}||^2$, 
where $N$ is the number of sub-models used in the cascade network, $\mathbf{S}_i$ is the output of sub-model $i$, $\mathbf{G}$ is the ground-truth image.

\section{Experiments}

\textbf{Evaluation metrics.} 
To evaluate the similarity between the original and the dewarped CT film image, we choose PSNR, SSIM~\cite{wang2004image}, and MS-SSIM~\cite{wang2003multiscale} as our evaluation metrics. 
In these experiments, to avoid the effects of image shifting, we estimate the \textbf{de-shifted} results on these metrics. 
In medical image restoration part, we add chi-square tests on radiomics features to check if it is practical.

\textbf{Training Details.}
For training, firstly, input images are resized to $256\times256$ and we set Adam~\cite{DBLP:journals/corr/KingmaB14} as the optimizer, learning rate to 0.001 and batch size to 16. All maps and images are linearly scaled to the range $[-1,1]$.

\subsection{The performance of FIReNet}

We test DewarpNet \cite{das2019dewarpnet} and our method on our CTFilm20k dataset.
DewarpNet directly generates a backward map $\mathbf{B}$, while we generate a UV map, which is more robust for film restoration.
Fig.\ref{fig:visual_comparison} which demonstrate a difficult example and two real photos and results in Table~\ref{table:compare} show our advantage over other methods. The improvement on the performance of evaluation in Table \ref{table:compare} confirms the benefit of using the UV map and our network architecture. 

\begin{table} 
    \begin{center}
    \begin{tabular}{lccc}
    \hline
    Methods & PSNR $\uparrow$  & SSIM $\uparrow$ & MS-SSIM $\uparrow$  \\
    \hline
    DewarpNet~\cite{das2019dewarpnet} &16.98  &0.4501 &0.6879\\
    FIReNet & 25.30 & 0.8621 & 0.9523 \\
    FIReNet(de-shifted) & 25.60 & 0.8715 & 0.9564\\
    \hline
    \end{tabular}
    \end{center}
    \caption{\textbf{Quantitative comparison between our FIReNet and  DewarpNet~\cite{das2019dewarpnet}.} Methods are evaluated by PSNR, SSIM and MS-SSIM. For all of these metrics, higher value means better. ``FIReNet" and ``FIReNet(de-shifted)" are the evaluated with normal backward map, and ``de-shifted" backward map (eq.\ref{eq:transLoss}), respectively. Our method performs better than DewarpNet~\cite{das2019dewarpnet}.}
\label{table:compare}
\end{table} 

\begin{figure}
\begin{minipage}[b]{.30\linewidth}
    \centering
    \begin{tabular}{ccc}
    \multicolumn{3}{c}{Metrics on DICOM} \\
    \hline
    Output & PSNR   & SSIM  \\
    \hline
    $\mathbf{I}_d$ & 5.23  & 0.593  \\
    $\mathbf{I}_r$ & 27.1  & 0.942 \\ 
    \hline
    \end{tabular}
\label{table:recovery}

\end{minipage}
\begin{minipage}[b]{.70\linewidth}
    \centering
    \begin{subfigure}{.32\textwidth}
    \centering
    \includegraphics[width=\linewidth]{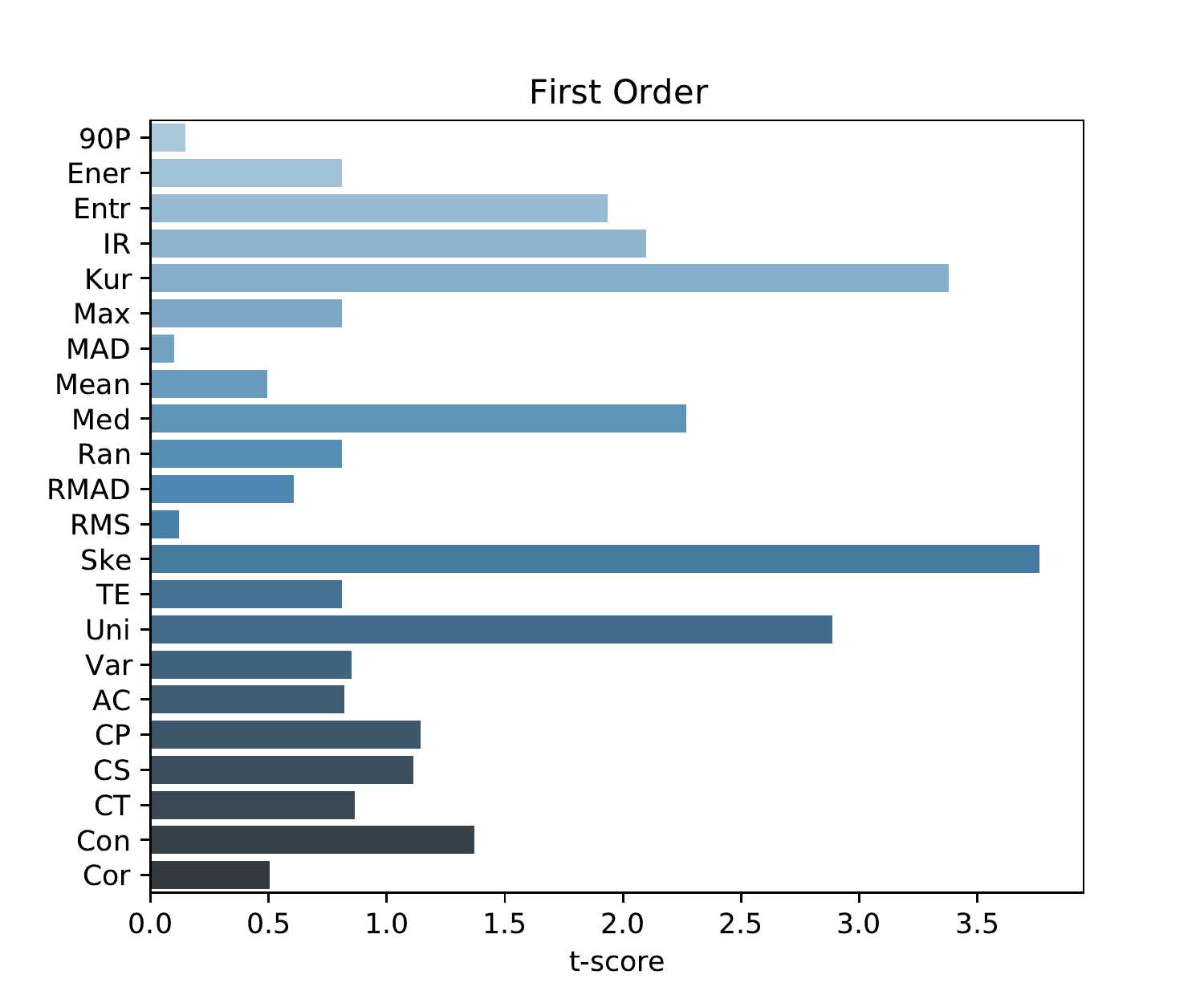}
    \end{subfigure}
    \begin{subfigure}{.32\textwidth}
    \centering
    \includegraphics[width=\linewidth]{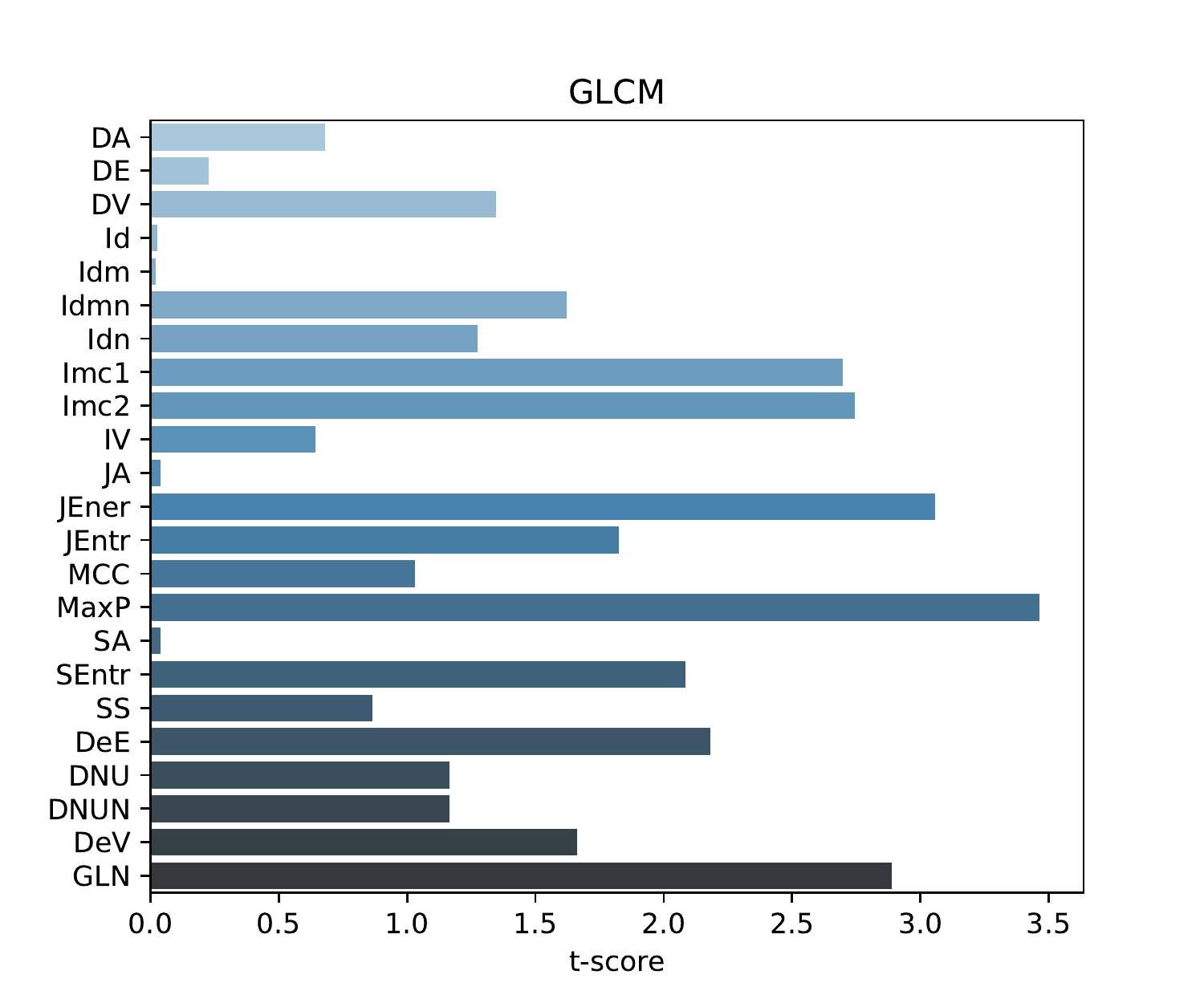}
    \end{subfigure}
    \begin{subfigure}{.32\textwidth}
    \centering
    \includegraphics[width=\linewidth]{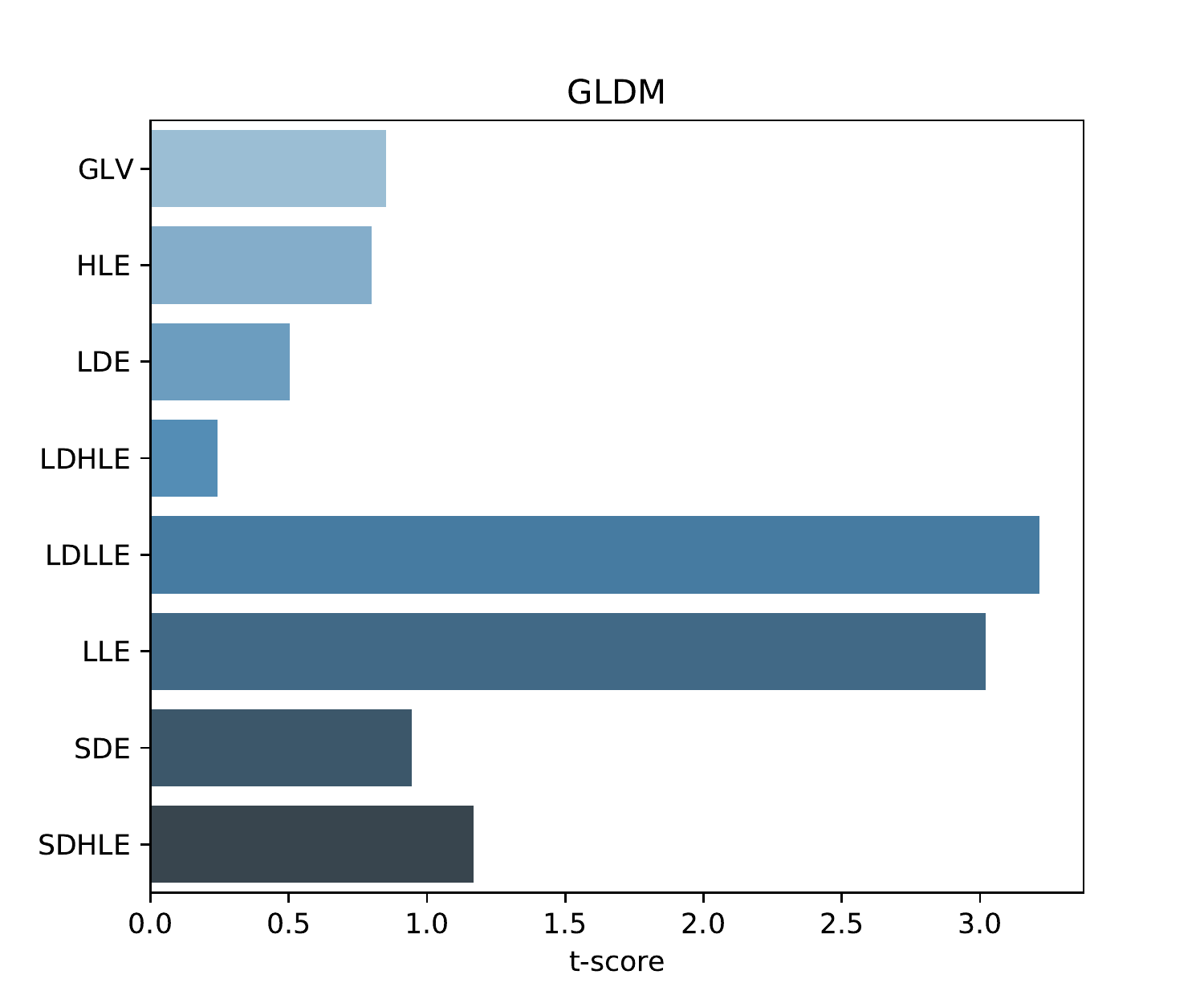}
    \end{subfigure}
    \label{fig:barfig}
\end{minipage}
    \caption{Metrics (Table) and t-score (Barplot) between our predictions and medical images (CT) as ground-truth. }
\label{fig:tscore}
\end{figure}

Besides, we test the medical image restoration module on our paired data including dewarped images from DewarpNet and the original DICOM files which is used for standard medical analysis. We directly input the dewarped images and predict the DICOM slice. The table in Fig.~\ref{fig:tscore} shows the huge gap between photos and DICOM files. Fig~\ref{fig:tscore} shows that the recovered DICOM files from our medical image restoration module passed the Chi-square test and is practical in the extent of reliable accuracy. 
We first suppose $H_0$: the distribution of prediction is identical to the ground-truth distribution, and we sample 101 predictions so the DOF (degrees of freedom) is 100. The significance level $\alpha$ is set to 0.05, 0.01, 0.001, and $t_{100}=1.660, 2.364,3.390$. 
If $t>t \frac{\alpha}{2}, n-1$, the assumption is rejected and the sample mean is different with population mean, vice versa. 
As shown in Fig.\ref{fig:tscore}, t-scores of almost features are smaller than the threshold. 67 in 101 are smaller than $t^{0.05}_{100}$, 83 are smaller than $t^{0.01}_{100}$, 98 are smaller than $t^{0.001}_{100}$.

\subsection{Ablation Study}

To further analyze our method, we conduct an ablation study to characterize the roles of each module. Results are shown in Table \ref{table:ablation}.
\textbf{UV map}: UV map is the key part of our method to restore the image. We compare our model which generates UV maps for restoration, with the baseline model which generates backward maps directly for restoration. According to Table~\ref{table:ablation}, we can find that UV map outperforms the backward map (25.16dB vs 23.76dB in PSNR).
\textbf{Multi-map}: 
We study the comparison between using only a 3D coordinate map and using multiple maps that include 3D coordinate, normal, and depth maps. 
From Table \ref{table:ablation}, we can find an improvement on these metrics (from 25.16dB to 25.51dB in PSNR).
\textbf{Deformation module}: We compare the performance of networks with and without the deformation module. The performance is improved from 25.51dB to 25.60dB in PSNR.

\begin{table}
\begin{center}
\begin{tabular}{ccc|ccc}
\hline
UV  & MM  & Deform &   PSNR $\uparrow$ & SSIM $\uparrow$ & MS-SSIM $\uparrow$\\
\hline
&& & 23.76 & 0.8450 &0.9446\\
 \checkmark & & &  25.16 & 0.8594 & 0.9515 \\
 \checkmark & \checkmark & & 25.51 & 0.8691 & 0.9553 \\
 \checkmark & \checkmark & \checkmark & 25.60 & 0.8715 & 0.9564 \\
\hline
\end{tabular}
\end{center}
\caption{\textbf{Ablation study.} UV: using UV maps. MM: generating multiple maps. Deform: adding deformation module.}
\label{table:ablation}
\end{table}

\section{Conclusions and Future Work}
In this work, we proposed our FIReNet, a novel dewarping network for CT film recovery. Our approach is robust to film content, reflection, and background and is insensitive to slight image shifting. Through extensive experiments, we validated the advantages of our approach, which outperforms the previous approaches. Moreover, we contribute the CTFilm20K dataset, which is the first large dataset for CT film dewarping, and full of various types of maps and annotations. However, there still exists some limitations like limited types of CT film. We will collect more data and validate our method to real scenarios.

\bibliographystyle{splncs04}
\bibliography{reference}

\end{document}